# On the Free Energy of the Flexomagnetoelectric Interactions

**B. M. Tanygin**[a]

[a] Kyiv Taras Shevchenko National University, Radiophysics Faculty, Glushkov av.2, build.5, Kyiv, Ukraine, MSP 01601

*Corresponding author:* B.M. Tanygin, 64 Vladimirskaya str., Taras Shevchenko Kyiv National University, Radiophysics Faculty. MSP 01601, Kyiv, Ukraine.

*E-mail*: b.m.tanygin@gmail.com

*Phone*: +380-68-394-05-52

**Abstract.**

It was shown that free energy density of the local flexomagnetoelectric effect is determined by the four phenomenological constants in case of the cubic (hexoctahedral) crystal. The well-known single-constant Lifshitz invariant term is correct only when fixed electric polarization induces the inhomogeneity of the magnetization. Proposed phenomenological theory was applied to the magnetic domain walls. The domain wall structure has been investigated in details. The four-constant phenomenological theory conforms to the symmetry based predictions (*V.G. Bar'yakhtar et al., 1984*). The proposed experimental verification of the four-constant flexomagnetoelectric phenomenology is a detection of the shift of the Néel domain walls under the strong homogeneous electric field.

**Keywords:** magnetoelectric effect, free energy, symmetry, domain wall.

## 1. Introduction

The magnetoelectric coupling mechanism [1] is of considerable interest to fundamentals of condensed matter physics and for the applications in the novel multifunctional devices [2]. Different phenomena correspond to the cases of the coupling with homogeneous [3-7] and inhomogeneous [8-13] magnetization. The latter type of phenomena exists in the magnetic domain walls [8,12-17] and magnetic vortexes [18,19]. This types of the magnetoelectric interaction have been described by the Lifshitz invariant-like single-constant coupling term $P_i\left(M_j\nabla_k M_n - M_n\nabla_k M_j\right)$ [11,17]. It was called as flexomagnetoelectric (FME) interaction [20-22]. The FME interaction is the particular case [8] of more general inhomogeneous magnetoelectric coupling [23] that can be determined by the spatial derivates of the polarization.

The FME effects have been observed in different experiments: switching the helicity with electric field in orthorhombic manganites $\mathrm{RMnO_3}$ $(\mathrm{R = Dy, Tb})$ [24,25] and $\mathrm{MnWO_4}$ [26], the motion of domain walls in the gradient electric field in $\mathrm{(BiLu)_3(FeGa)_5O_{12}}$ films [14,15], incommensurate spin-density-wave state observed in the ferroelectric $\mathrm{BiFeO_3}$ [27]. The indirect evidence of the FME effect can be seen in the enhancement of electromagnetooptical effect (i.e. electric field induced Faraday rotation) in the vicinity of domain wall observed in yttrium-ferrite-garnet films [13]. Also, the FME effect can account for the anomalies in dielectric function observed in the re-entrant phase of $\mathrm{HoMnO_3}$ [28].

The symmetry based analysis [12] of the FME generation of the electric polarization was compared [29] with the equilibrium polarization obtained from the single-constant FME term. These results are identical for the transverse component of **P** but different for the domain wall plane normal component $P_z$. The magnetic symmetry approach allows component $P_z$ for all 64 magnetic point groups. The single-constant expression suppresses component $P_z$ for all magnetic domain walls. Consequently, the fundamental symmetry based theory [12] does not coincide with the Lifshitz invariant based phenomenology results for all components.

The Lifshitz invariant was initially obtained [30] by removing the total spatial derivatives $\nabla_k M_j M_n$ from the free energy density expression because they do not affect the final solution. This rule is correct in case when unknown function of the variational problem is $\mathbf{M}(\mathbf{r})$. Let us call this type of the variational problem as magnetization problem that requires the polarization distribution $\mathbf{P}(\mathbf{r})$ to be like a fixed parameter. The example is the electric field induced helicoidal magnetic structure in the orthorhombic antiferromagnetic [31]. Also the magnetic vortex (antivortex) [18] and magnetic domain walls [14] micromagnetic structure control by the electric fields have been well described in the scope of the magnetization problem. The magnetization problem statement is to find the shift of the magnetic domain wall (changes of the micromagnetic structure) in dependence on the external electric field. The fixed parameter $\mathbf{P}(\mathbf{r})$ for all these cases means that polarization $\mathbf{P_0}(\mathbf{r}) = \chi_e \mathbf{E}(\mathbf{r})$ is determined by the external field only. The additional requirement of the magnetization problem is that polarization $\mathbf{P}_{\boldsymbol{ME}}(\mathbf{r})$ [17] induced by the FME coupling should be substantially smaller than $\mathbf{P_0}(\mathbf{r})$.

In case when unknown function of the variational problem is $\mathbf{P}(\mathbf{r})$ (polarization problem) or both $\mathbf{M}(\mathbf{r})$ and $\mathbf{P}(\mathbf{r})$ (magnetization-polarization problem) the approach with removing $\nabla_k M_j M_n$ terms from the free energy is not correct because solution $\mathbf{P}_{\boldsymbol{ME}}(\mathbf{r})$ depends on these terms. Without removing of the total spatial derivatives from the free energy density of crystal the single-constant coupling term cannot be obtained from the symmetry of the cubic $m\bar{3}m$ crystal. Consequently, the single-constant phenomenology of the induced polarization $\mathbf{P}_{\boldsymbol{ME}}$ is not consistent. More general free energy expression containing terms $\nabla_k M_j M_n$ should be formulated. Building of such theory for the case of the cubic (hexoctahedral) crystal is the purpose of this work.

## 2. Phenomenological theory

The invariant of the FME interactions in the free energy of the crystal is given by general form [5,8,10] (here and hereinafter we use Einstein notation):

$$F_{ME} = \gamma_{ijkn} P_i M_j \nabla_k M_n \,, \tag{1}$$

where structure of the tensor $\gamma_{ijkn}$ is determined by the crystal symmetry. Here only the first in order of magnitude terms [12] are considered. Also, only short-range contribution [12] to the FME coupling was taken into account. If take into account that $\mathbf{M}$ is an axial time-odd vector, the $\mathbf{P}$ is a polar time-event vector, the action of operator $\nabla_k$ is the analog of the multiplication on the $k$-th component of the polar time-even vector [32] then it is possible to find invariants of the type (1) which are allowed by the magnetic symmetry of the cubic $m\bar{3}m$ crystal (group $m\bar{3}m1'$, where $1'$ is a time reversal transformation). The following invariants remain constant after action of any symmetry transformation of the group $m\bar{3}m1'$:

$$I_1 = P_i M_i \nabla_i M_i, \tag{2a}$$

$$I_2 = P_i M_j \nabla_i M_j (1 - \delta_{ij}), \tag{2b}$$

$$I_3 = P_i M_j \nabla_j M_i (1 - \delta_{ij}), \tag{2c}$$

$$I_4 = P_i M_i \nabla_j M_j (1 - \delta_{ij}), \tag{2d}$$

The symmetry transformations of the group $m\bar{3}m1'$ do not transform one of these invariants into other. Thus, they must be present in the free energy of the crystal with different phenomenological constants.

The magnetization problem is a case of the unknown function $\mathbf{M}(\mathbf{r})$ and fixed electric polarization. In this case the invariants (2a-2b) are total space derivatives of unknown functions. Consequently, these terms can be removed from the free energy as well as the total space derivates transform into zero in the Euler-Lagrange equation. After removing of total space derivate the terms (2c-2d) convert into one term with a single constant. The free energy of FME interactions takes the form [11]:

$$F_{ME} = \gamma_0 \mathbf{P}[\mathbf{M}(\nabla\mathbf{M}) - (\mathbf{M}\nabla)\mathbf{M}] \tag{3}$$

The polarization problem is case of the unknown function $\mathbf{P}(\mathbf{r})$ and fixed magnetization (the $\mathbf{M}(\mathbf{r})$ presents in the problem like a parameter). The magnetization-polarization problem is case of the unknown functions $\mathbf{M}(\mathbf{r})$ and $\mathbf{P}(\mathbf{r})$. The magnetization-polarization problem is most general and physically complete approach. The total space derivatives of the $\mathbf{M}(\mathbf{r})$ influence final solution of the variational problem because corresponding terms cannot be removed from all Euler-Lagrange equations. It means that single-constant Lifshitz invariant-like term (3) is not correct for the description of the free energy in scope of the polarization and magnetization-polarization problem. The expression (3) is only approximate description [17] of the phenomena of these problems.

The general (correct for magnetization, polarization or magnetization-polarization problem) FME term of the cubic $m\bar{3}m$ crystal is four-constant term:

$$F_{ME} = \gamma_1 I_1 + \gamma_2 I_2 + \gamma_3 I_3 + \gamma_4 I_4 \tag{4}$$

Taking into account (2), we obtain:

$$F_{ME} = \gamma_1 I_1 + \gamma_2 \left[\frac{(\mathbf{P}\nabla)\mathbf{M}^2}{2} - I_1\right] + \gamma_3[\mathbf{P}(\mathbf{M}\nabla)\mathbf{M} - I_1] + \gamma_4[(\mathbf{PM})(\nabla\mathbf{M}) - I_1] \tag{5}$$

It is useful to introduce new phenomenological constants:

$$\tilde{\gamma}_1 = \gamma_1/2 - \gamma_2 - \gamma_3 - \gamma_4 \tag{6a}$$

$$\tilde{\gamma}_2 = \gamma_2/2 \tag{6b}$$

$$\tilde{\gamma}_3 = \gamma_3 \tag{6c}$$

$$\tilde{\gamma}_4 = \gamma_4 \tag{6d}$$

Using these constants the final FME term is given by:

$$F_{ME} = \tilde{\gamma}_1 P_i \nabla_i M_i^2 + \tilde{\gamma}_2 (\mathbf{P}\nabla)\mathbf{M}^2 + \tilde{\gamma}_3 \mathbf{P}(\mathbf{M}\nabla)\mathbf{M} + \tilde{\gamma}_4 (\mathbf{PM})(\nabla\mathbf{M}) \tag{7}$$

The second term of (7) describes the type of FME interactions which exist only if magnetization magnitude is spatially modulated [33]. It was stated [12] that such type of the FME coupling is possible in the vicinity of the Curie point or when elementary magnetic cell contains more than one magnetic atom. The fourth term of (7) exists only in micromagnetic structures with stray field. It is necessary to compare expressions (3) and (7). The material constants of these expressions are related:

$$\gamma_0 = \frac{\tilde{\gamma}_3 - \tilde{\gamma}_4}{2} \tag{8}$$

The types of FME interactions describing by the material parameters $\tilde{\gamma}_1$ and $\tilde{\gamma}_2$ do not affect the equilibrium electronic and magnetic crystal structure in case of the magnetization problems.

The free energy (7) determines the direction of the internally induced polarization $\mathbf{P}_{ME}$. The magnitude and direction of this polarization is determined in the similar way as for the dielectric in presence of the external field $\mathbf{P}_{ME} = \chi_e \mathbf{E}_{ME}$ [17], where $\chi_e$ is the scalar electric susceptibility of the cubic material and $\mathbf{E}_{ME} = -\partial F_{ME}/\partial \mathbf{P}$ is the effective field of the FME coupling. The total polarization is:

$$\mathbf{P} = \chi_e[\mathbf{E} + \tilde{\gamma}_1 \mathbf{e}_i \nabla_i (M_i^2) + \tilde{\gamma}_2 \nabla(\mathbf{M}^2) + \tilde{\gamma}_3 (\mathbf{M}\nabla)\mathbf{M} + \tilde{\gamma}_4 \mathbf{M}(\nabla\mathbf{M})] \tag{9}$$

**3. Results and discussion**

The electric polarization that appears in the magnetic domain walls must be considered in scope of the polarization or magnetization-polarization problem. The magnetization-polarization problem is most exact physical model and it can be solved by the micromagnetic simulation (with variation of both **M** and **P**) with term (7) in the free energy of crystal. Let us analyze the significantly simpler model: the polarization problem. The polarization problem is a valid model for almost all magnetically ordered materials. The FME effect is novel phenomenon [19] but it is theoretically possible (symmetry predictions) in any magnetically material [12,29,31]. The root cause if this irrelevance is that FME effective field is small, i.e. usually it does not affect the magnetization distribution. Example of the experimental investigation that can be well described by the polarization problem is the FME effect in the magnetic domain walls of the iron garnet film [14]. In the framework of the one-dimensional model of the magnetic domain wall without presence of the external field the expression (9) obtains the form:

$$\mathbf{P} = \chi_e \left[ \tilde{\gamma}_1 \mathbf{e}_z \frac{\partial M_z^2}{\partial z} + \tilde{\gamma}_2 \mathbf{e}_z \frac{\partial \mathbf{M}^2}{\partial z} + \tilde{\gamma}_3 M_z \frac{\partial \mathbf{M}}{\partial z} + \tilde{\gamma}_4 \mathbf{M} \frac{\partial M_z}{\partial z} \right] \tag{10}$$

where the term $\tilde{\gamma}_2 \partial \mathbf{M}^2 / \partial z$ can be obviously removed far from the Curie point. Component of the polarization are given by the expressions:

$$P_x = \chi_e \left[ \tilde{\gamma}_3 M_z \frac{\partial M_x}{\partial z} + \tilde{\gamma}_4 M_x \frac{\partial M_z}{\partial z} \right] \tag{11a}$$

$$P_y = \chi_e \left[ \tilde{\gamma}_3 M_z \frac{\partial M_y}{\partial z} + \tilde{\gamma}_4 M_y \frac{\partial M_z}{\partial z} \right] \tag{11b}$$

$$P_z = \chi_e \left[ \left( \tilde{\gamma}_1 + \frac{\tilde{\gamma}_3 + \tilde{\gamma}_4}{2} \right) \frac{\partial M_z^2}{\partial z} + \tilde{\gamma}_2 \frac{\partial \mathbf{M}^2}{\partial z} \right] \tag{11c}$$

The most significant feature of the symmetry analysis of the FME interaction is a possibility to determine types of the component of the function $\mathbf{P}(z)$ for all 64 magnetic point groups of the domain walls [29]. Only 57 groups remain in the cubic crystal [34]. The magnetic point symmetry based predictions for the components $P_x(z)$, $P_y(z)$ and $P_z(z)$ was formulated [12,29] in terms of even or odd function, or sum of odd and even functions. Such predictions have been compared [29] with ones obtained from the existing phenomenological expressions, i.e. the Lifshitz invariant-like (3). In case of the transverse components the symmetry based predictions were identical with them. Predicted by the symmetry existing permanently component $P_z(z)$ was not allowed by FME phenomenology (3). Consequently, the analysis of the phenomenological background for the inducing of the longitudinal polarization component is urgent in scope of the present work.

The phenomenological results (11a-b) give the identical types of the transverse components as in the work [29]. The expression (11c) shows that $P_z(z)$ is odd in case when $\mathbf{M}(z)$ or $M_z(z)$ are either odd or even. All these functions can have (F) type only in the same time. The term $\tilde{\gamma}_2\,\partial\mathbf{M}^2/\,\partial\mathrm{z}$ is important only for domain walls which appear near the point of the phase transition [33,35] (table 1, types III-IV). The magnetic point group of the domain wall is derived from the crystal symmetry and cannot exceed it [36]. The magnetic point group of the domain wall is determined by the domain wall plane orientation, crystal sample orientation. Influence of the crystal sample free surfaces on the symmetry of the domain wall has been investigated [34] which automatically describe influence of the depolarization field (as a part of the $\mathbf{E}$ in (9)) on the domain wall structure. The different magnetic point groups with the same type of the magnetization distribution can produce different polarization because these domain walls appear in different media (table 1).

As a rule, all domain walls with type III are achiral (according to the chirality definition [29]). Details of the relation between magnetic point group of the domain wall and its micromagnetic structure were presented for different cases in the [23,29,34,36]. The Bloch-like domain walls (type I) have weak or zero (if $|\mathbf{M}|$ is exactly equal to the saturation magnetization $M_S$) induced longitudinal polarization. They are the Bloch domain wall with center [35] ($2'_\mathrm{x}2_\mathrm{y}2'_\mathrm{z}$), Bloch domain wall without center ($2'_\mathrm{z}$) and Bloch-like 0° domain wall [36-38] ($2'_\mathrm{z}/\mathrm{m}'_\mathrm{z}$). The Néel-like domain walls (type II) have induced longitudinal polarization with order of magnitude $\chi_e M_S^2[\tilde{\gamma}_1+(\tilde{\gamma}_3+\tilde{\gamma}_4)/2]/\delta_0$, where $\delta_0$ is width of the domain wall. They are the Néel domain wall without center ($\mathrm{m}'_\mathrm{x}$), the charged [33] Néel domain wall with center ($\mathrm{m}'_\mathrm{x}\mathrm{m}'_\mathrm{z}2_\mathrm{y}$), the Néel-like 0° domain wall ($2'_\mathrm{x}/\mathrm{m}'_\mathrm{x}$), the non-charged Néel domain wall with center ($\mathrm{m}_\mathrm{z}\mathrm{m}'_\mathrm{y}2'_\mathrm{x}$), etc.

Thus, new phenomenological approach show almost complete coincidence with the symmetry based predictions [29]. The 52 of total 57 magnetic point groups show full coincidence for the three components of the polarization predicted by symmetry and expression (10). Remain 5 groups (bold in the column III of table 1) describe symmetry of domain walls with single magnetization component and two polarization components. The full conformity between phenomenology and symmetry based predictions appear if we take into account additional types of the inhomogeneous magnetoelectric interactions with spatial derivatives of the polarization [23]. The main peculiarity of these phenomenological results is presence of the odd function in the longitudinal polarization component (coupled electric charge in the domain wall volume). Consequently, experimental verification of the free energy expression (7) can be provided using the homogeneous electric field control of the micromagnetic structure in the experimental setup similar as described in the [14].

**Table 1.** Magnetic point groups of the domain walls. Types of the magnetization distribution and induced polarization distribution are specified*.

| **I** | **II** | **III** | **IV** | |
|---|---|---|---|---|
| $M_z = 0, \frac{\partial \mathbf{M}^2}{\partial z} = 0$ | $M_z \neq 0, \frac{\partial \mathbf{M}^2}{\partial z} = 0$ | $M_z = 0, \frac{\partial \mathbf{M}^2}{\partial z} \neq 0$ | $M_z \neq 0, \frac{\partial \mathbf{M}^2}{\partial z} \neq 0$ | |
| $2'_z/m'_z$:<br>[SS0] → [00A] | $m_z m'_y 2'_x$:<br>[A0S] → [S0A] | $m_x m_z m'_y$:<br>[A00] → [00A] | $4_z/m'_z m'_x m'_{xy}$:<br>[00A] → [00A] | $\bar{3}_z m'_x$:<br>[00S] → [00A] |
| $2'_x 2_y 2'_z$:<br>[AS0] → [00A] | $m'_x m'_z 2_y$:<br>[0SA] → [0SA] | $m_y m'_x m'_z$:<br>[0S0] → [00A] | $4_z/m_z m'_x m'_{xy}$:<br>[00S] → [00A] | $3_z 2_x$:<br>[00A] → [00A] |
| $2'_z$:<br>[FF0] → [00F] | $2'_x/m'_x$:<br>[0SS] → [0AA] | $\mathbf{m_y m'_z 2'_x}$:<br>[0S0] → [S0A] | $4_z m'_x m'_{xy}$:<br>[00F] → [00F] | $3_z 2'_x$:<br>[00S] → [00A] |
| | $m'_x$:<br>[0FF] → [0FF] | $m'_y m_x 2'_z$:<br>[F00] → [00F] | $\bar{4}'_z 2_x m'_{xy}$:<br>[00A] → [00A] | $\bar{3}'_z$:<br>[00A] → [00A] |
| | $m'_z$:<br>[SSA] → [SSA] | $\mathbf{m_x m_z 2_y}$:<br>[A00] → [0SA] | $\bar{4}_z 2'_x m_{xy}'$:<br>[00S] → [00A] | $\bar{3}_z$:<br>[00S] → [00A] |
| | $m_z$:<br>[AAS] → [SSA] | $\mathbf{2'_x/m_x}$:<br>[A00] → [0AA] | $4_z 2_x 2_{xy}$:<br>[00A] → [00A] | $3_z$:<br>[00F] → [00F] |
| | $2'_x$:<br>[ASS] → [SSA] | $2'_z/m_z$:<br>[AA0] → [00A] | $4_z 2'_x 2'_y$:<br>[00S] → [00A] | $m_z m'_x m'_y$:<br>[00S] → [00A] |
| | $2_y$:<br>[ASA] → [ASA] | $\mathbf{2_y/m_y}$:<br>[0S0] → [A0A] | $4_z/m'_z$:<br>[00A] → [00A] | $m'_x m'_y m'_z$:<br>[00A] → [00A] |
| | $\bar{1}$:<br>[SSS] → [AAA] | $\mathbf{m_y}$:<br>[F00] → [0FF] | $4_z/m_z$:<br>[00S] → [00A] | $m'_x m'_y 2_z$:<br>[00F] → [00F] |
| | 1:<br>[FFF] → [FFF] | | $\bar{4}'_z$:<br>[00A] → [00A] | $2_y/m_y'$:<br>[A0A] → [A0A] |
| | | | $\bar{4}_z$:<br>[00S] → [00A] | $2_z/m'_z$:<br>[00A] → [00A] |
| | | | $4_z$:<br>[00F] → [00F] | $2_z/m_z$:<br>[00S] → [00A] |
| | | | $3_z m'_x$:<br>[00F] → [00F] | $2_x 2_y 2_z$:<br>[00A] → [00A] |
| | | | $\bar{3}'_z m'_x$:<br>[00A] → [00A] | $2_z 2'_x 2'_y$:<br>[00S] → [00A] |
| | | | | $2_z$:<br>[00F] → [00F] |
| | | | | $\bar{1}'$:<br>[AAA] → [AAA] |

* $[T_x\ T_y\ T_z]$ marks the type of the magnetization (left) and polarization (right) distribution: even (S) or odd function (A), or sum of odd and even functions (F).

## 4. Conclusions

Thus, the free energy density of the local flexomagnetoelectric effect is determined by the four phenomenological constants. The single-constant Lifshitz invariant term is valid only for particular cases. The four-constant phenomenological theory has a full conformity with the symmetry based predictions. The experimental verification of the four-constant flexomagnetoelectric phenomenology is a shift of the magnetic domain walls using the strong homogeneous electric field.

**Acknowledgements**

I would like to express my sincere gratitude to my supervisor Prof. V.F. Kovalenko for his outstanding guidance and to my wife D.M. Tanygina for spell checking.

## References

[1] W. Eerenstein, N. D. Mathur, J. F. Scott, Nature 442 (2006) 759-765.

[2] M. Bibes, A. Barthelemy, Nat. Mater. 7 (2008) 425–426.

[3] D. Khomskii, Physics 2 (2009) 20.

[4] L. Landay, E. Lifshitz, Electrodynamics of Continuous Media, Pergamon Press, Oxford, 1965.

[5] G. Smolenskii, I. Chupis, Usp. Fiz. Nauk 137 (1982) 415.

[6] I. Dzyaloshinskii, Zh. Eksp. Teor. Fiz., 37 (1959) 881.

[7] N. Neronova and N. Belov, Dokl. Akad. Nauk SSSR 120 (1959) 556.

[8] V.G. Bar'yakhtar, V.A. L'vov, D.A. Yablonskiy, JETP Lett. 37 (12) (1983) 673.

[9] G.A. Smolenskii, I. Chupis, Sov. Phys. Usp. 25(7) (1982) 475.

[10] I. M. Vitebskii, D. A. Yablonski, Sov. Phys. Solid State 24 (1982) 1435.

[11] A. Sparavigna, A. Strigazzi, A. Zvezdin, Phys. Rev. B 50 (1994) 2953.

[12] V.G. Bar’yakhtar, V.A. L’vov, D.A. Yablonskiy, Theory of electric polarization of domain boundaries in magnetically ordered crystals, in: A.M. Prokhorov, A.S. Prokhorov (Eds.), Problems in Solid-State Physics, Mir Publishers, Moscow1984, pp. 56–80 (Chapter 2).

[13] V.E. Koronovskyy, S.M. Ryabchenko, V.F. Kovalenko, Phys. Rev. B 71 (2005) 172402.

[14] A. S. Logginov et al., JETP Lett. 86 (2007) 115; Appl. Phys. Lett. 93 (2008) 182510.

[15] A. P. Pyatakov, et al., EPL 93 (2011) 17001.

[16] A.P. Pyatakov, et al., J. Phys.: Conf. Ser. 200 (2010) 032059.

[17] M. Mostovoy, Phys. Rev. Lett. 96 (2006) 067601.

[18] A. P. Pyatakov, G. A. Meshkov and A. S. Logginov, Moscow Univ. Phys. Bull. 65 (4) (2010) 329-331.

[19] A. S. Logginov et al., J. Magn. Magn. Mater 310 (2007) 2569–2571.

[20] A.P. Pyatakov, A.K. Zvezdin, Eur. Phys. J. B 71 (2009) 419.

[21] A. K. Zvezdin, A. A. Mukhin, JETP Lett. 89 (7) (2009) 328.

[22] A.P. Pyatakov, A.K. Zvezdin, Phys. Stat. Sol. (b) 246 (8) (2009) 1956.

[23] B.M. Tanygin, IOP Conf. Ser.: Mater. Sci. Eng. 15 (2010) 012073.

[24] E. Milov, et al., JETP Lett. 85 (2007) 503.

[25] Y. Yamasaki, et al., Phys. Rev. Lett. 98 (2007) 147204.

[26] T. Finger, et al., Phys. Rev. B 81 (2010) 054430.

[27] A. Kadomtseva, A. Zvezdin, Y. Popov, A. Pyatakov, and G. Vorob’ev, JETP Lett. 79 (2004) 571.

[28] T. Lottermoser and M. Fiebig, Phys. Rev. B 70 (2004) 220407(R).

[29] B.M. Tanygin, J. Magn. Magn. Mater 323 (5) (2011) 616-619.

[30] E.M. Lifschitz, JETP 11 (1941) 253.

[31] V.G. Bar’yakhtar, D.A. Yablonskiy, Sov. Phys. Solid State 24 (1982) 2522.

[32] L. D. Landau and E. M. Lifshitz, Statistical Physics, Addison-Wesley, Reading, Mass., 1958.

[33] A. Hubert, R. Shafer, Magnetic Domains. The Analysis of Magnetic Microstructures, Springer, Berlin, 1998.

[34] B.M. Tanygin, O.V. Tychko, Acta Physic. Pol. A. 117 (1) (2010) 214-216.

[35] V. Bar'yakhtar, V. L’vov, D. Yablonsky, Sov. Phys. JETP 60(5) (1984) 1072-1080.

[36] B.M. Tanygin, O.V. Tychko, Phys. B: Condens. Matter 404 (21) (2009) 4018-4022.

[37] L. Heyderman, H. Niedoba, H. Gupta and I. Puchalska, J. Magn. Magn. Mater 96 (1991) 125.

[38] R. Vakhitov, A. Yumaguzin, J. Magn. Magn. Mater. 52 (2000) 215-216.